\documentclass[pre,nofootinbib,showpacs,floatfix,twocolumn]{revtex4}
\usepackage{graphicx}

\newcommand{\LiI}{\mbox{Li\hspace{2bp}{\sc i}}}

\begin{document}

\title{Investigation of electric fields in plasmas under pulsed currents}

\author{K.~Tsigutkin}
 \altaffiliation[Present address: ]{Department of Physics University of
 California, 366 LeConte Hall MC 7300, Berkeley, CA 94720-7300}
 \email{tsigutkin@berkeley.edu}
\author{R.~Doron}
\author{E.~Stambulchik}
\author{V.~Bernshtam}
\author{Y.~Maron}
 \affiliation{Faculty of Physics, Weizmann Institute of Science, Rehovot 76100,
              Israel}
\author{A.~Fruchtman}
 \affiliation{Sciences Department, Holon Institute of Technology, Holon 58102,
              Israel}
\author{R.~J.~Commisso}
 \affiliation{Pulsed Power Physics Branch, Plasma Physics Division, Naval
              Research Laboratory, Washington, DC 20375-5346, USA}

\date{\today}

\begin{abstract}
Electric fields in a plasma that conducts a high-current pulse are
measured as a function of time and space. The experiment is
performed using a coaxial configuration, in which a current rising
to 160~kA in 100~ns is conducted through a plasma that prefills
the region between two coaxial electrodes. The electric field is
determined using laser spectroscopy and line-shape analysis.
Plasma doping allows for 3D spatially resolved measurements. The
measured peak magnitude and propagation velocity of the electric
field is found to match those of the Hall electric field, inferred
from the magnetic-field front propagation measured previously.
\end{abstract}


\maketitle

\section{Introduction}
\label{introduction} For a long time, laboratory experiments and
observations in space have been stirring up a dispute concerning the
mechanisms of the magnetic-field evolution in plasma and the
associated plasma dynamics. Estimates of the plasma collisionality
in plasma opening switches (POS), for example, indicate that the
magnetic field diffusion should be small while the plasma pushing by
the magnetic-field pressure should be dominant. However,
magnetic-field penetration has been observed in POS
experiments~\cite{Weber}, and, moreover, it is found to be
significantly faster than expected from
diffusion~\cite{Spitalnik,PRL,B-fieldRon,Dynamics}. The
magnetic-field penetration was also found to be higher than, or of a
comparable velocity to the plasma pushing. In an early
experiment~\cite{Spitalnik} a fast magnetic-field
penetration into a plasma where the ions are nearly immobile was observed. This
was explained by the generation of an electric Hall field based on
the electron magnetohydrodynamics
theory~\cite{Kingsep1,Gordeev1,Fruchtman,FruchtmanPRA}. The relative
importance of the two governing processes, i.e., the plasma pushing
and the magnetic-field penetration, was further analyzed in a series
of additional experiments with different plasma parameters, in which
comprehensive measurements of spatial distributions of the magnetic
field, of the ion velocity, and of the electron  density in the
plasma were performed. These detailed measurements, performed in
multi-ion species plasma, have revealed that the fast magnetic field
penetration into the plasma is accompanied by a specular reflection
of the light-ion plasma and a slow pushing of the heavy-ion plasma
penetrated by the field~\cite{B-fieldRon, Dynamics}, resulting in
ion-species separation~\cite{PRL}.

The electric field plays a crucial role in the interaction of the
magnetic field with the plasma. The electric field transfers
energy to the current-carrying plasma through the Joule heating,
the electric field that is generated by the space-charge
separation due to the magnetic field accelerates the ions, and the
(Hall) electric field may induce the magnetic-field penetration
into the plasma. However, previous studies of the electric fields
in plasmas under application of pulsed currents (e.g.,
Refs.~\cite{Weingarten_turbulence,B-fieldRon,Dynamics}) lacked the
temporal and spatial resolutions and, thus, only provided an
averaged view.

In this paper we describe an application of 3D-resolved spectroscopy to
investigate the time evolution of electric fields in plasmas under high-current
pulses.  Our method is based on laser spectroscopy combined with plasma doping
by lithium, which allows for electric field measurements from the lithium emission-line shape.
In the following sections we first discuss the
physics underlying the diagnostic method and the measurements of the initial
conditions of the plasma prior to the application of the current. Then,
measurements of the electric field in the plasma under a current pulse are
described, followed by a discussion on the relation between the measured electric field, the magnetic
field, and the plasma dynamics. It is found that the
measured electric field magnitude is consistent with the Hall electric field calculated on the basis of
the magnetic field and the plasma density measured in a separate study.

\section{The diagnostic method}
\label{method}%
The Stark effect has proven to be of a great utility for
non-intrusive measurements of the frequency and amplitude of local
electric fields in plasmas, as first suggested by Baranger and
Mozer~\cite{Baranger} and shortly later experimentally verified by
Kunze and Griem~\cite{Kunze-Griem}.

The advantages of the lithium atomic system for electric fields
measurements have been previously
demonstrated~\cite{Rebhan-Kunze,Hildebrandt,Rebhan}. Lithium was
employed for measurements of electric fields in dilute plasmas of
gas discharges~\cite{Burrell,Kawasaki} and in the vacuum gap of a
high-voltage diode~\cite{Knyazev}. In our experiment, high spatial
and temporal resolutions in measuring the electric fields in
relatively dense ($\sim 10^{14}~cm^{-3}$) plasma are achieved. At
such densities, extracting the information from the spectral
line-profiles becomes more difficult due to the increased Stark
broadening and the faster ionization processes that reduce the
neutral lithium line emission.

The E-field diagnostic is based on the analysis of the line-shape of a
forbidden transition and its intensity relative to an allowed
transition. The neutral-lithium atomic system is most suitable for
applying the method since the 2p-4f dipole-forbidden transition
(4601.5~\AA) and the 2p-4d allowed transition (4603~\AA) are very close in
wavelength, and therefore, both lines can be recorded
simultaneously. This is particularly useful for investigating a
highly transient plasma, such as the present one, where a reliable
measurement requires obtaining the information on the E-field in a
single discharge. In the presence of an electric field, a
configuration mixing of the 4d and 4f levels occurs, making possible
an electric dipole transition from the 4f level to the 2p level.
Since the 4d-4f mixing coefficient depends on the electric field
amplitude, the intensity of the forbidden transition can serve as a
measure of the electric field strength. The mixing coefficient is
larger for smaller energy-level difference, thus the small energy
separation between the 4f and the 4d levels ($\approx 5~\text{cm}^{-1}$) makes these
transitions in lithium suitable for measurements of relatively low electric
field. The time dependence of the electric field magnitude and direction
also affects the spectral line-profiles,
and thus, using a detailed line-shape analysis allows for obtaining additional
information on the frequency spectrum of the E-fields.
In the case of a harmonic perturbation with frequency $\omega$,
a dipole-forbidden transition is split into two satellites,
displaced from the unperturbed position by $\pm \hbar \omega$~\cite{Baranger}.
Generally, for a non-harmonic perturbation the linewidth is roughly $2 \hbar
\left< \omega \right>$,
where $\left< \omega \right>$
is the typical field frequency. Therefore, taking into consideration also the contributions of the Doppler and instrumental broadenings to the total line widths,
it is possible to conclude that for the present case of Li I spectrum,
the forbidden line would be unresolved from the strong allowed line for E-field frequencies
above $\approx~20~\text{GHz}$. Our detailed modeling confirms this.

Under the relevant plasma conditions the populations of the 4d and 4f levels are
insufficient to produce strong spectral lines required for accurate
measurements. Laser-induced-fluorescence (LIF) technique is employed to increase
the 4d and 4f level populations by the laser pumping. Pumping of these levels
directly from the ground state (2s) is inefficient since the 2s--4d and 2s--4f
transitions are dipole-forbidden. Instead, the laser emission is employed for
pumping the 4p level from the ground state, and electron-collisional transfer
from the 4p level leads to the increase of the 4d and 4f levels populations. The
knowledge of the relative populations of the 4d and 4f levels is important for
proper interpretation of the relative line intensities. Due to the small energy
separation between the 4d and 4f levels, it is expected that the
electron-collision processes between these two levels dominate the relative
populations of these levels, leading to a population distribution in accordance
to the level statistical weights. This is verified by collisional-radiative
calculations~\cite{CR}. The collisional excitations and de-excitations from the
pumped 4p level lead to an increased population also of the levels of the
neighboring n=3 configuration. The resulting rise in the line intensities is
advantageous, since different lines are used to obtain simultaneously both the
Stark and the Doppler contributions to the line shapes. For example, due to the
different sensitivity of the 4d and 3d states to the Stark effect, a comparison
of the 2p--3d and 2p--4d line shapes allows for inferring the Doppler
broadening and the rather small Stark-broadening contribution to the 2p--4d
shape.
\begin{figure}
 \includegraphics*[width=\columnwidth] {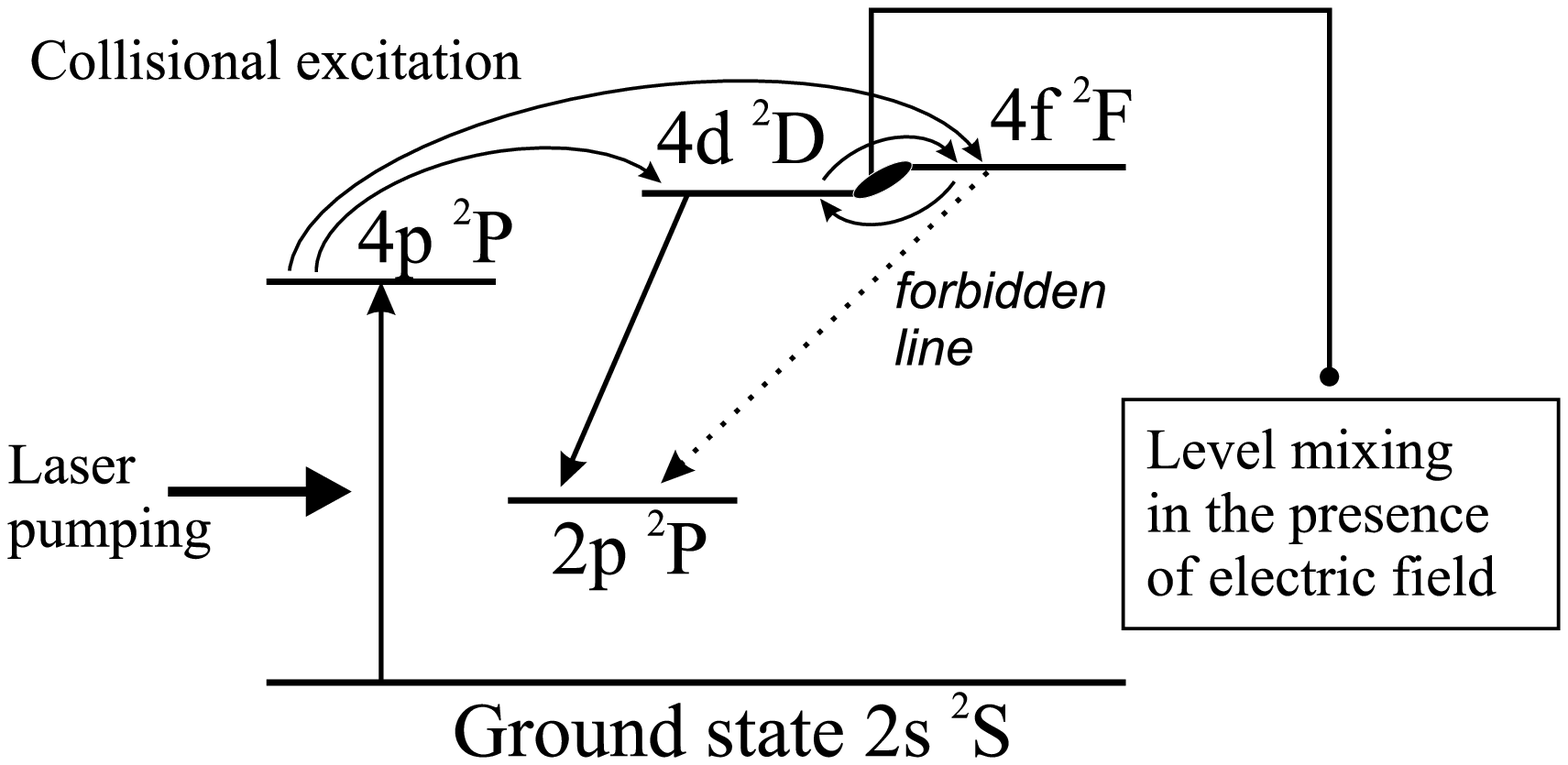}
 \caption{The scheme of laser-driven excitations of the \LiI\ levels.}
 \label{levels}
\end{figure}

\begin{figure}
 \includegraphics*[width=\columnwidth] {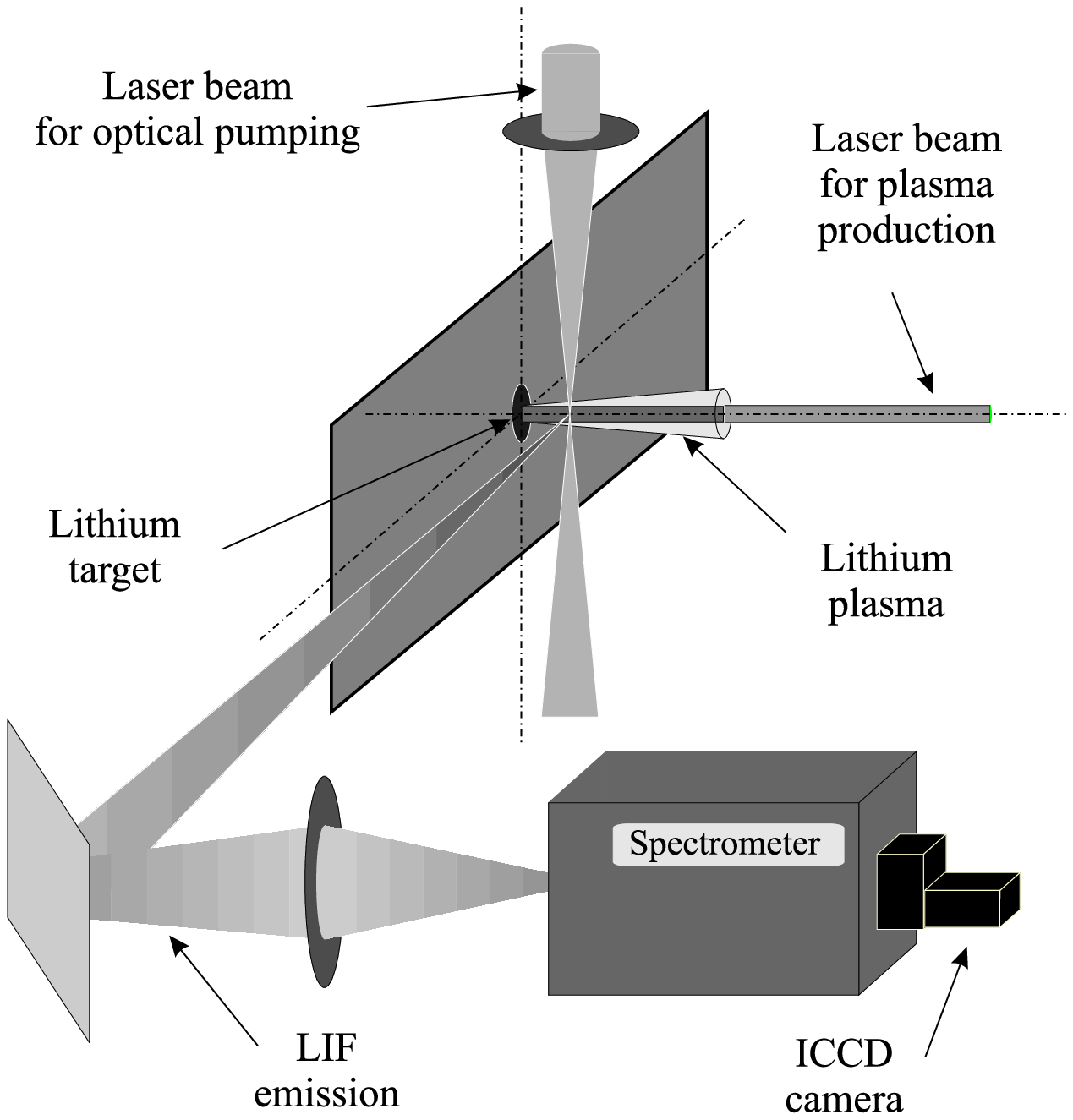}
 \caption{Application of LIF for the electric field measurements.}
 \label{doping}
\end{figure}

In the present experiments a lithium beam, produced by applying a
Nd:YAG laser pulse onto a solid lithium target, is injected into the
plasma region of interest. Another laser beam, which is used for the
pumping, intersects the lithium beam at the point of interest to
produce the LIF emission (see Fig.~\ref{doping}). The lithium beam
density is so selected that it introduces a minimal disturbance in
the region of diagnostics, while, on the other hand, is sufficient
for yielding a satisfactory intensity of the desired transitions.

For the line-shape modeling, the method described in
Ref.~\cite{Evgeny} was used. This method is capable of calculating
line shapes under the simultaneous effects of the plasma micro- and
macro-fields (both electric and magnetic ones). The forbidden
transition from the $4f$ to the $2p$ level of \LiI\ is caused mostly
by the 4f--4d level mixing. The mixing of the 2p and 4f levels with
other levels is negligible since the energy separations between
these levels and 2p and 4f are significantly larger than the 4d--4f
separation. Therefore, the 4f--4d energy separation is crucial for
an accurate modeling. We have measured its value
independently~\cite{4d-4f} in order to resolve the ambiguity in
light of the large spread in the values of this quantity in the
available data sources.

\section{Experimental setup}
\label{setup} 
\begin{figure}
 \includegraphics*[width=\columnwidth] {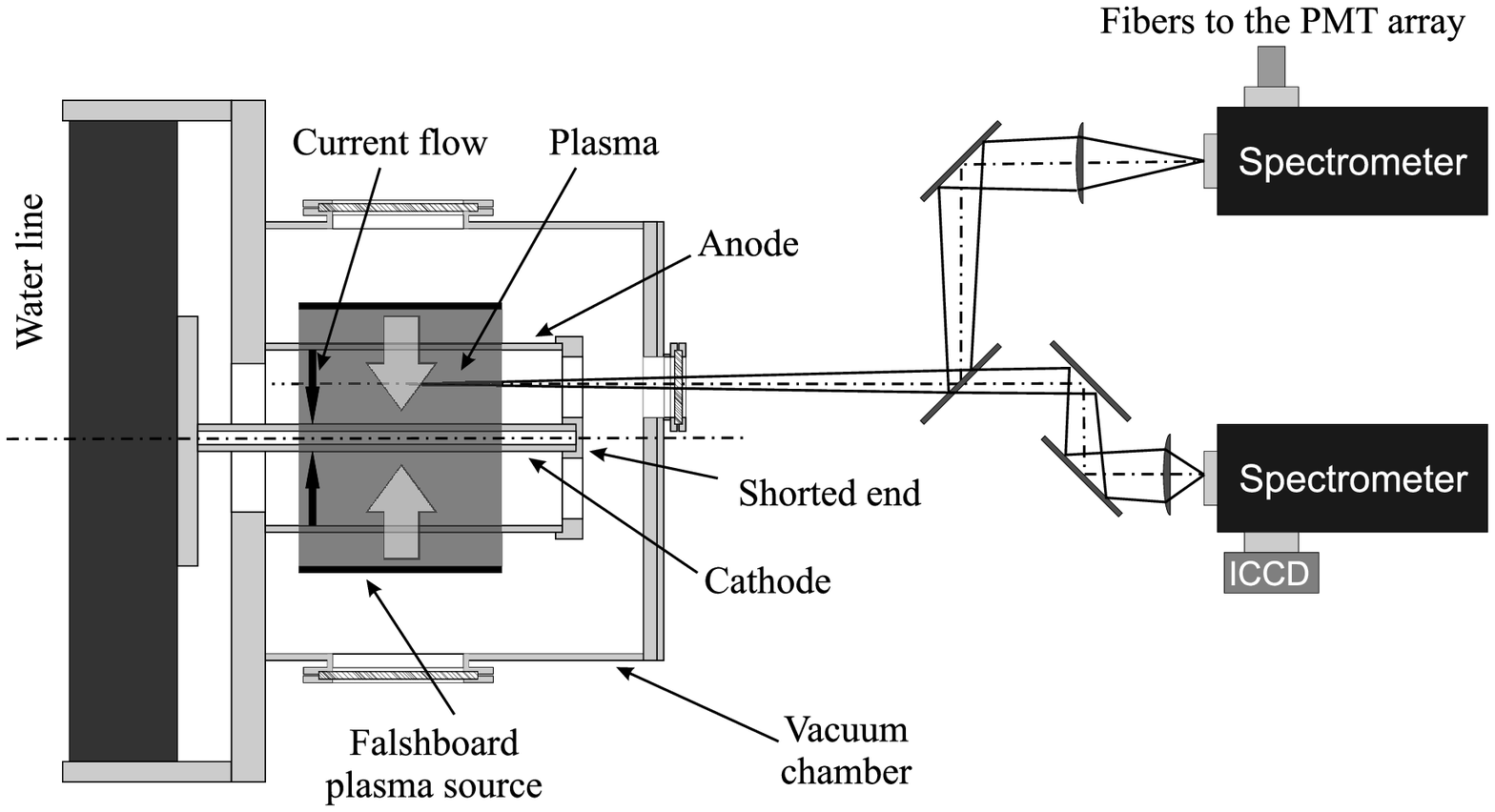}
 \caption{The layout of the coaxial transmission line and the diagnostic system.}
 \label{posdiag}
\end{figure}

In this experiment we employ a coaxial plasma
configuration, shown in Fig.~\ref{posdiag}, in which plasma is
initially injected to fill up the volume between the two coaxial
electrodes. The cathode and anode diameters are 3.8 and 8.9 cm,
respectively, and the plasma is injected over an axial length of 10
cm. A current pulse applied at one side of the coaxial line
propagates through the plasma until the current channel reaches the
vacuum section of the transmission line that is followed by a
shorted end. This experimental configuration is commonly referred
to as plasma opening switch (see e.g.~\cite{POS ref} and reference
therein) and has been recently used for studying the interaction of
the plasma with the propagating magnetic
field~\cite{PRL,B-fieldRon,Dynamics,Rami}.

The plasma prefill in the present experiment is produced by a
flashboard plasma source and is injected inward through the
transparent anode to fill up the A-K gap (see Fig.~\ref{posdiag}).
The flashboard plasma with an electron density $n_e = 2\times 10^{14}
~\text{cm}^{-3}$ and electron temperature $T_e \simeq 5$~eV,
almost uniform across the A-K gap, is mainly composed of protons and
carbon ions (mainly C III-V). Since the heavier carbon ions flow at
lower velocities than the protons, the carbon/proton relative
abundance decreases towards the cathode.

The coaxial line is powered by an LC-water-line Marx generator. A
current pulse of $160~\text{kA}$ with a rise time of
$100~\text{ns}$ is then driven through the coaxial line. The
upstream and downstream currents are measured by Rogowski coils
placed at the output of the water-line and at the shorted end.

\begin{figure}
 \includegraphics*[width=\columnwidth] {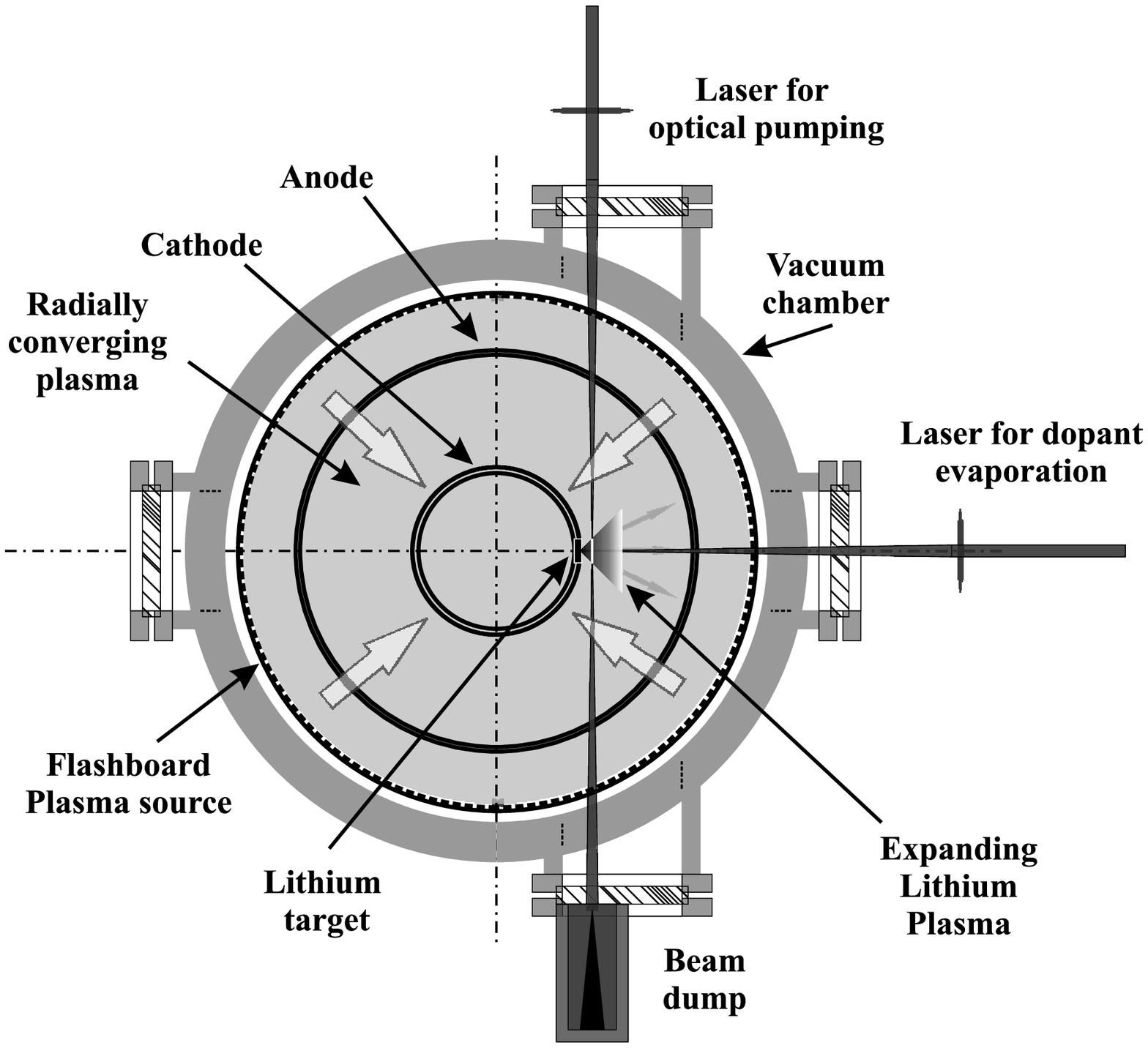}
 \caption{The cross section of the coaxial chamber, showing the dopant excitation
          scheme. Observations are performed in the direction perpendicular to
          the figure plane.}
 \label{POScross}
\end{figure}

In the diagnostic technique here employed, the flashboard prefill plasma is doped with
lithium using a laser-produced lithium beam (see Fig.~\ref{POScross}). For the
generation of the lithium beam, a Nd:YAG laser pulse (6 ns) is focused onto a
surface of a metallic lithium target attached to one of the coaxial electrodes. The
laser intensity was $5\times 10^7~\text{W/cm}^2$ on the target surface. The
particle velocities in the laser-produced plasma plume were found to be $(1\div
10)\times10^5~\text{cm/s}$. At the distance of $2.5~\text{mm}$ from the target,
at which the electric field measurements are performed, the density of the
lithium beam was $\simeq 7\times 10^{13}~\text{cm}^{-3}$ with an ionization
degree of $\approx 65\%$, i.e., the electron density was $\simeq 4\times
10^{13}~\text{cm}^{-3}$.  The electron temperature was $\simeq 0.5~\text{eV}$
(see Sec.~\ref{optim}). For the photopumping of the \LiI\
4p state, a tunable dye laser is used. This laser is pumped by an additional
Nd:YAG laser pulse of  15~ns duration. In order to obtain the wavelength
required for the photopumping ($\lambda_{2s-4p}=2741.2~\mathring{A}$), a
wavelength extender for the generation of the second harmonic of the dye laser
pulse is employed. Thus, the dye laser is tuned to generate a pulse at
$\lambda_{dye} = 5482.4$~\AA\ (using a solution of \texttt{Fluorescein~548} dye
in methanol). The energy achieved at 2741.2~\AA\ is $\simeq$1~mJ. The
laser spectral width is 0.2~\AA.

For measurements where recording of accurate spectral line profiles are important,
experimental conditions for which the Doppler broadening is relatively small
are preferable. The Doppler broadening that affects the line shapes in these
measurements results from the lithium atom velocities in the axial (\textit{Z}) direction
(along the line of sight). In order to minimize this effect it is required to
use a sufficiently long time delay between the laser pulse for target evaporation and the current pulse
application. The reason is that the lithium atoms ejected from the target
surface due to the laser pulse move away from the surface with velocities
varying along the plasma column produced. Therefore, slow atoms reach the
measurement region with longer delays. Generally, lower velocities in the
propagation direction normal to the target surface are correlated with lower velocities in the transverse
directions, which enabled us to achieve a satisfactorily low Doppler broadening
in the axial direction. In the present experiment we used a time delay of 450~ns
between the laser pulse employed for the target evaporation and the laser pulse employed for photopumping.
Selecting longer time-delays decreases the amount of the atoms
due to the lithium ionization processes that take place in the prefill plasma, thereby, reducing
the signal-to-noise ratio. Therefore, the optimal delay was determined according
to the requirement of minimal Doppler broadening but high enough signal
intensity.

The diagnostic system consists of two 1-m UV-visible spectrometers
equipped with 2400~grooves/mm gratings (see Fig.~\ref{posdiag}).
The output of one spectrometer is collected by an optical-fiber
array and transmitted to 10 photomultiplier tubes that allow the
recording of the line-profile time-dependence (temporal resolution of
7~ns), while the output of the second spectrometer is recorded by
a gated (down to 5~ns) intensified charge-coupled device (ICCD) camera,
allowing for
recording single gated broadband spectra. The spectral resolutions
of both systems are $\simeq 0.2$~\AA\ in the spectral range used. Since the induced fluorescence
originates from the region along the pumping-laser path, observing
the emitted radiation perpendicular to the pumping-beam direction
allows for measurements with a high spatial resolution in 3
dimensions. In this case the spatial resolution along the line of
sight is determined by the width of the pumping laser beam, which
is $\simeq 1.5~\text{mm}$, whereas the resolution in the plane
perpendicular to the line of sight is chosen by the spectrometer slit
dimension (sub-mm scale). A detailed discussion of the spectral calibration and
the determination of the instrumental response of the system can
be found in Ref.~\cite{4d-4f}.

\section{Results}
\label{results}
The measurements are performed in three stages. In the first stage we
characterize the laser-produced lithium plasma used as a dopant. In principle,
the dopant plasma could perturb the local conditions of the main prefill plasma,
introducing uncertainties in the measurements. In order to minimize this
perturbation, the dopant density must be kept lower than that of the main
plasma. However, lowering the lithium density results in weakening the
spectral lines of interest, thereby reducing the signal-to-noise ratio that is
crucial for resolving the spectral profiles required for the determination of the
electric field. To optimize the \LiI\ dopant density, measurements of the lithium density, ionization degree, and
expansion velocity were first performed with no plasma
prefill.

In the second stage, with the optimal configuration of the lithium doping, the
electric fields are measured in the plasma prefill (still
without the application of the current pulse). These experiments yield the
initial level of the electric field present in the plasma prior to the
application of the current pulse. We refer to this field as a ``background''
electric field.

Finally, we measure the evolution of the electric fields generated
during the conduction of the pulsed current.

\subsection{Diagnostics of the lithium dopant beam}
\label{optim} Details of the measurements aimed at the diagnostics
of the laser-produced lithium beam will be published
separately~\cite{f-lines}. Here, we briefly describe the main
results. Measurements were performed to determine the temporal
evolution of the plasma parameters in the lithium beam at different
distances from the target for various laser intensities in the range of
$10^7$ - $10^8~\text{W/cm}^2$

\begin{figure}
 \includegraphics*[width=\columnwidth] {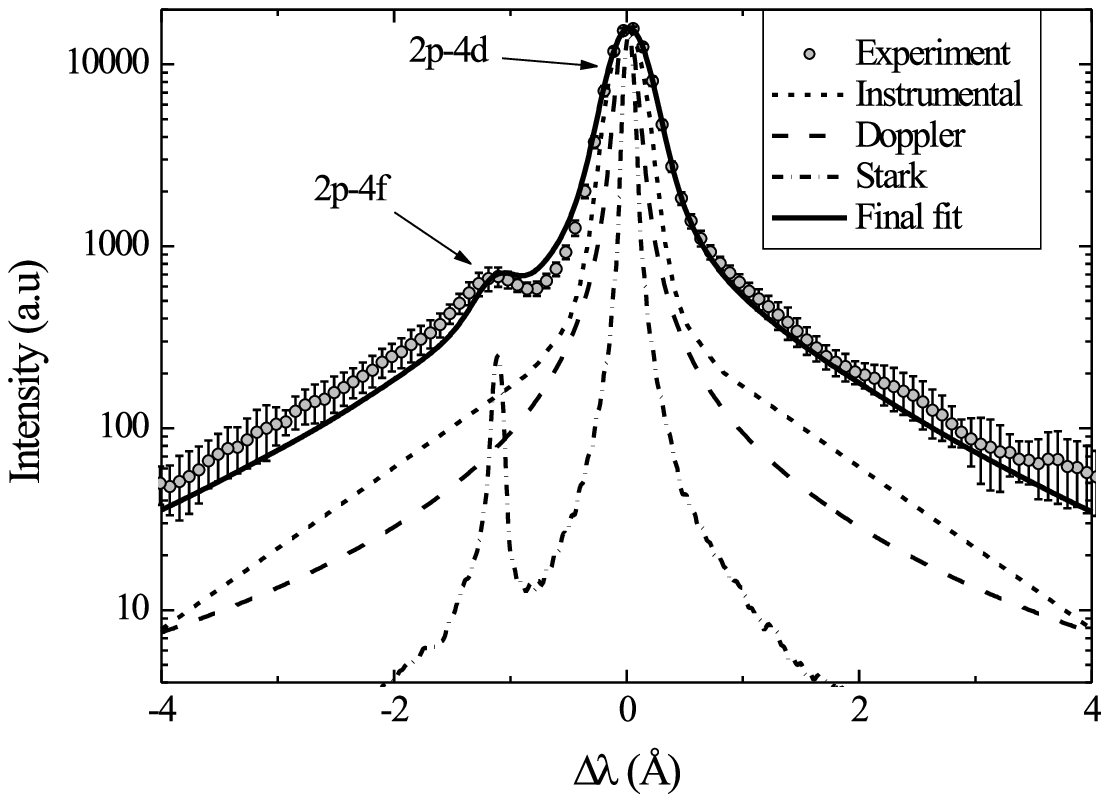}
 \caption{Measured and calculated \LiI\ 2p--4d line
          shapes including the 2p--4f forbidden line.}
 \label{profile_las}
\end{figure}

A typical \LiI\ spectrum in the wavelength range of interest (4603~\AA), obtained at a distance of
2.5~mm from the target, is presented in Fig.~\ref{profile_las}. It is recorded by
the ICCD camera with a time-gate of 10~ns correlated with
the peak of the LIF signal. The figure shows the measured and simulated line shapes of
the allowed 2p-4d and the nearby forbidden 2p-4f transitions.
The line shape simulation, similar to that described in Ref.~\cite{4d-4f},
show the individual contribution of each of the three
broadening mechanisms: the Stark effect, Doppler, and instrumental.

The Doppler broadening can be estimated from the \LiI\ velocities
obtained from time-of-flight measurements. For the present
measurements we used a time delay of 450 ns between the evaporating
laser pulse and the photopumping laser pulse. At a distance of
2.5~mm from the electrode this delay corresponds to a velocity of
$5.5\times 10^{5}~\text{cm/s}$. Since the \LiI\ plume
expands into a large solid angle, the transverse velocities are of
the order of the longitudinal velocities. Indeed, a Doppler
broadening of 2~eV, consistent with the velocities mentioned above,
is obtained by deconvolving the known instrumental broadening
function from the total line shape of the \LiI\ 2p--3d transition
(6104~\AA) that is insensitive to the Stark effect under the present
conditions.

Subsequently, the Stark-broadened spectrum for various values of
the electron density $n_e$ is calculated, and convolved with the
instrumental and the Doppler broadening functions. For these
calculations, we used $T_e=0.5~\text{eV}$, adopted from the results of a
collisional-radiative modeling of the temporal behavior of the
\LiI\ line intensities (described in Ref.~\cite{thesis}).

From the line-shape calculations it is found that the measured
spectrum cannot be satisfactorily fitted for any electron density.
A good fit is rather obtained by assuming a combination of
$n_e=(4\pm 0.7)\times 10^{13}~\text{cm}^{-3}$ and low frequency
($\lesssim\omega_{pi}$) oscillations with an amplitude of $\approx
3~\text{kV/cm}$ (see Fig.~\ref{profile_las}). Although assuming a
higher electron density of $n_e\simeq 2\times
10^{14}~\text{cm}^{-3}$ (with no waves in the plasma) provides a
satisfactory fit to the total intensity ratio of the allowed line
to the forbidden line, the width of the 2p-4f line becomes
significantly larger due to the wide spectrum of the microfield
amplitudes under this assumption. This results in flattening of
the forbidden peak, which is in disagreement with the
observations.

\subsection{Effect of the plasma prefill injection on the \LiI\ line shapes}
\label{sec_FB} We now describe the experiments in which the
lithium beam is locally injected into the flashboard plasma that
prefills the inter-electrode gap (still, without applying the main
pulsed-current). The lithium target is attached to the cathode
(the inner electrode, see Fig.~\ref{POScross}). The plasma prefill
parameters were obtained in a previous work from emission
spectroscopy~\cite{FB_ns}. The plasma is found to have a nearly
uniform electron density across the inter-electrode gap with
$n_e\simeq 2\times 10^{14}~\text{cm}^{-3}$ and $T_e\simeq
5~\text{eV}$. Under these conditions the lithium atoms undergo
substantial ionization and the density of neutral lithium arriving
at the observation point is expected to be significantly lower
than when the dopant is injected into vacuum, leading to reduced
line intensities. Indeed, the intensity of the \LiI\ 4603-\AA\
line in the plasma prefill is found to be an order of magnitude
lower than when the lithium beam is expanding in vacuum.
Nevertheless, this intensity is still sufficient to provide a
signal-to-noise ratio that is adequate for resolving the line
profiles of interest. In these experiments the 2p-3d line is not
useful for measuring the Doppler broadening since its upper level
is no longer populated from the decay of the $n=4$\ levels, as the
latter are significantly depopulated through collisional
ionization due to the higher prefill plasma density. Instead, here
we use the 2s-2p line at 6708~\AA\ that is also insensitive to the
Stark effect. This line exhibits somewhat larger Doppler width
than in the case of expansion into vacuum, corresponding to a
temperature of $\simeq 2.5~\text{eV}$.
\begin{figure}
 \includegraphics*[width=\columnwidth] {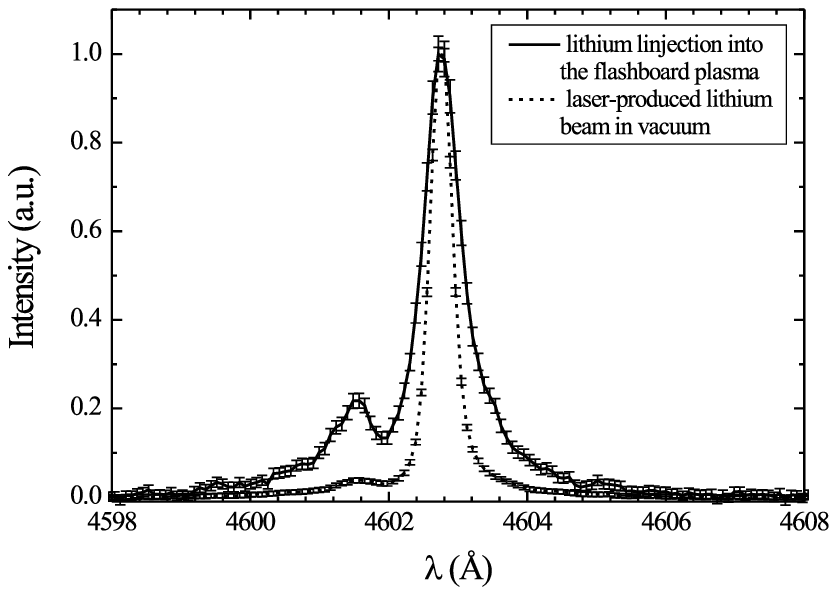}
 \caption{The spectrum of the 2p--4d and 2p--4f lines of \LiI\ recorded in experiments with and without the presence of the flashboard prefill plasma. The spectra are normalized to have the same peak intensity for the 2p--4d lines.}
 \label{FB_2p-4d}
\end{figure}

In Fig.~\ref{FB_2p-4d} we present a comparison of the
profiles of the allowed 2p--4d and the forbidden 2p--4f lines
emitted when the lithium beam is expanding in vacuum with those emitted with the flashboard plasma prefill.
It is clearly seen that in the presence of the flashboard plasma
the forbidden line has a substantially higher relative intensity
and both lines are relatively broadened. Similarly to the previous
case, also here we find that electric microfields, arising
from the thermal electrons and ions, cannot explain the observations. The
best fit is obtained by considering a combination of micro
electric fields due to $n_e = (1\div4)\times
10^{14}~\text{cm}^{-3}$ and the presence of a low frequency
electric field of $8\pm 1~\text{kV/cm}$. We note that under these
higher density conditions, the line profiles are less sensitive to
the density, but the inferred density range agrees well with that
found in a previous work~\cite{FB_ns}. The simulations also give
an upper limit for the oscillation frequency of the
additional electric field, $\approx 10~\text{GHz}$. This upper
limit is of the order of $\omega_{pi}$ of the flashboard plasma,
indicating that these oscillations can possibly be of the
ion-acoustic type. Similar E-field intensities were measured in several positions,
particularly in the positions where the E-field measurements were performed during the current application, as described in the next paragraph.
We take this observed electric field of $8~\text{kV/cm}$ to be the upper
limit of the collective fields present in the plasma
prefill prior to the application of the high-current pulse; hereafter
referred to as the ``background'' electric field.
\begin{figure}
 \includegraphics*[width=\columnwidth] {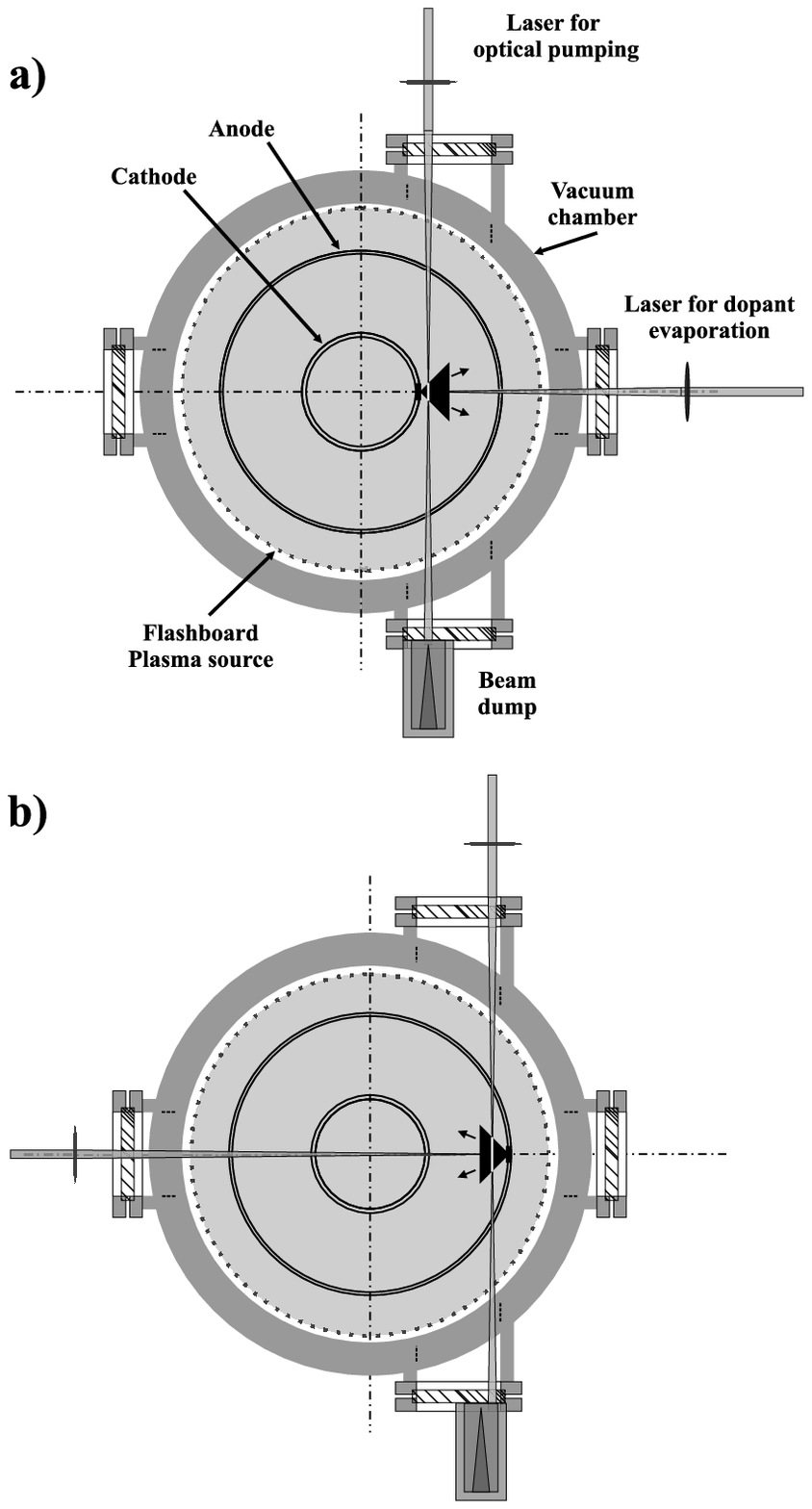}
 \caption{The two directions of the dopant injection. a) E-field measurement near the cathode.
 b) E-field measurement near the anode.}
 \label{beam_injection}
\end{figure}

\begin{figure}
 \includegraphics*[width=\columnwidth] {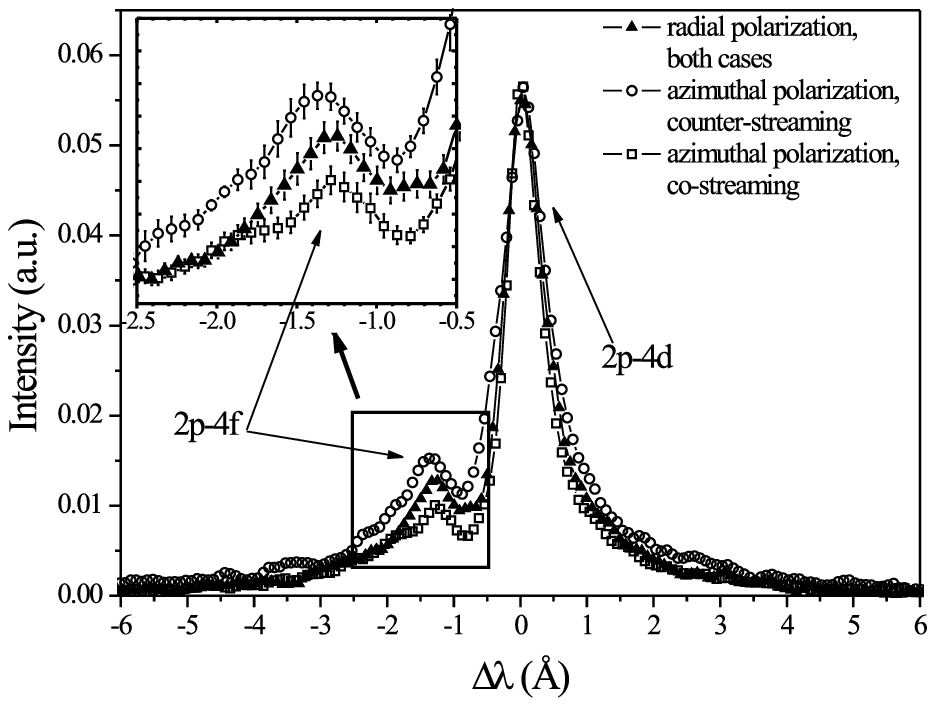}
 \caption{The profiles of the \LiI\ 2p--4d and 2p--4f transitions
          measured for two orthogonal polarizations for the two directions of
          the dopant injection.}
 \label{FB_pol}
\end{figure}

For performing measurements of electric fields also in the
vicinity of the anode, the lithium target is attached to the anode
and the evaporating laser beam is focused onto the target through
a port in the vacuum chamber at the opposite side of the anode.
The two different setups, allowing for measurements in the
vicinity of the cathode and the anode, are schematically described
in Fig.~\ref{beam_injection}. Since the flashboard plasma source
is located on an outside cylinder and its plasma flows radially
inward, attaching the dopant-lithium target to the anode results
in evaporated Li atoms flowing mostly in the same direction as the
flashboard plasma flow, whereas when the Li target is attached to
the cathode, the Li atoms flow opposite to the direction of the
flashboard plasma flow. The line profiles obtained near the anode
are compared to those obtained near the cathode. The comparison of
the line profiles is performed for azimuthal and radial
polarizations (the line of sight is along the axial direction).
Interestingly, as shown in Fig.~\ref{FB_pol}, the change of the
dopant injection direction with respect to the flashboard plasma
flow is found to affect the line profiles only in the azimuthal
polarization. For the azimuthal polarization, the 2p--4f
forbidden-line intensity is noticeably higher for the case where
the dopant \LiI\ atoms are injected opposite to the direction of
the flow of the flashboard plasma (measurements near the cathode),
while for the radial polarization the line profiles are found to
be similar for the two cases of the dopant injection. The
explanation for this effect is not clear yet. For the E-field
measurements in the present study only the radial-polarization
line-emission is used, for which no effect of the switching of the
lithium injection direction on the line shapes is observed.

\subsection{Electric field measurements in the plasma under the high-current pulse}
\label{sec_POS} 
\begin{figure}
 \includegraphics*[width=\columnwidth] {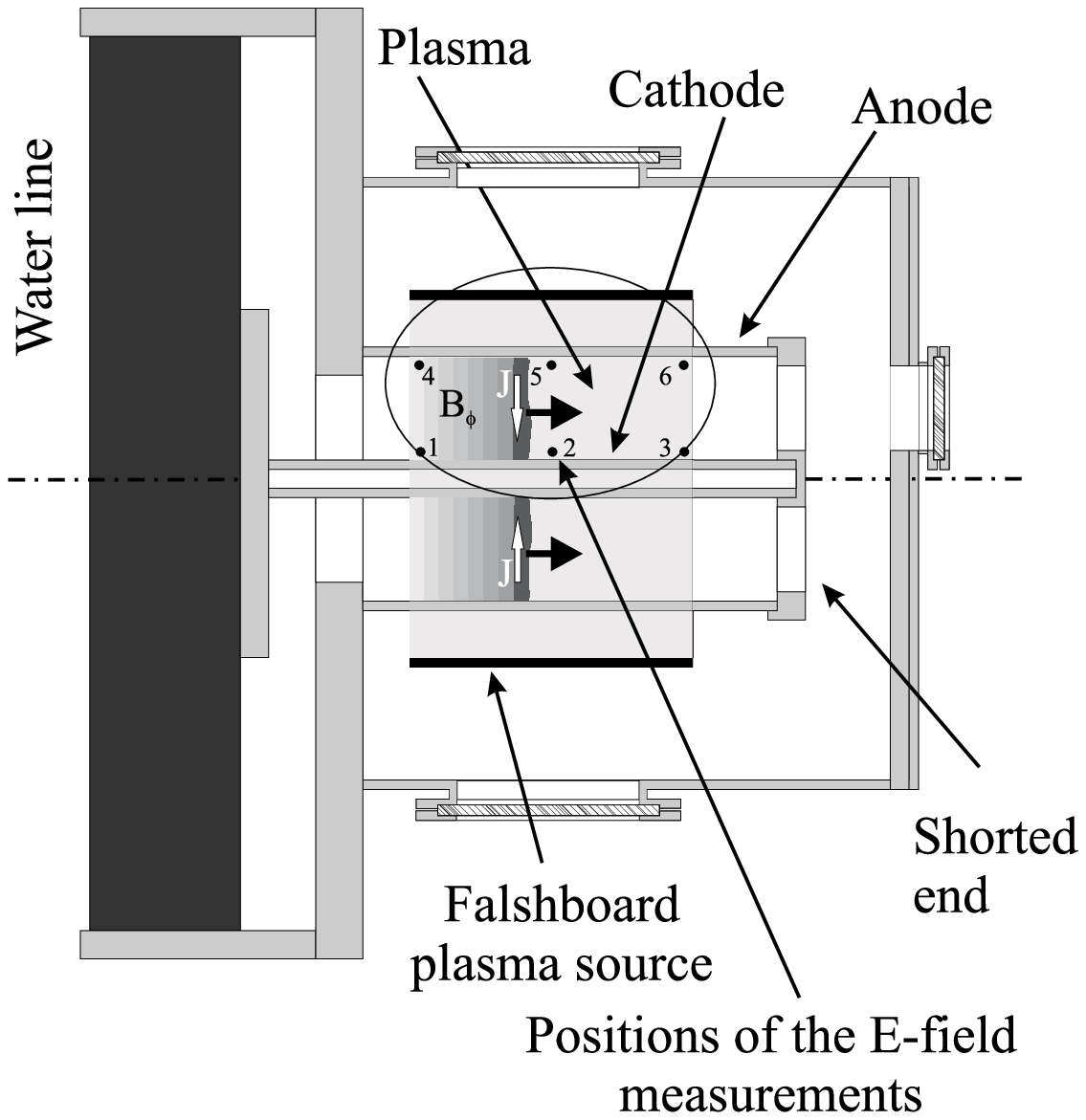}
 \caption{Positions for the E-field measurements during the current application.
          The positions are located symmetrically in the A-K gap having an axial
          separation of $35~\text{mm}$. The total length of the plasma is $\approx
          10~\text{cm}$.}
 \label{E_POS_setup}
\end{figure}

Knowledge of the effect of the prefill plasma on
the dopant lithium line shapes and the corresponding electric
fields allows for investigating the electric fields formed during
the flow of the pulsed-current in the plasma. The measurements are
performed at a distance of 2.5~mm from the electrodes (either the
cathode or the anode) inside the inter-electrode gap at different
axial positions. The field of view of the optical system covers a
region between 2.2 and 2.8~mm from the electrodes. We note that
estimates of the maximal sheath size give a much smaller distance
from the electrodes, based on the known system inductance and the
measured current. The electric field measured is thus in the
plasma and not in the sheath. The arrangement of the measurement
positions is shown in Fig.~\ref{E_POS_setup}. The measurement
positions are located symmetrically near the anode and near the
cathode separated axially by 35~mm. The positions \textbf{1} and
\textbf{4} in the plasma are located at a distance of $\approx
1.5$~cm from the generator-side plasma boundary. The boundary
position is defined as the position beyond which (towards the
generator) $n_e$ drops by more than a factor of 3 relative to that
in the main part of the plasma. Due to the plasma expansion
towards the cathode the plasma boundary is slightly farther from
the measurement positions near the cathode than near the anode.
This setup allows for determining the velocity of the electric-field
axial propagation at the two radial positions (near the anode and of
the cathode). The temporal evolution of the electric field at each
position is obtained by performing consecutive experiments, varying
the time delay between the application of the current pulse and the
application of the photo-pumping laser. The temporal resolution of
the measurements (10~ns) is determined by the time-gate of the ICCD
camera. In order to cover the entire duration in which the current
flows through the plasma, the E-field is obtained at a minimum of 6
consecutive time intervals in each position. For overcoming the
shot-to-shot irreproducibility, the results at each time are
averaged over several discharges.
\begin{figure}
 \includegraphics*[width=\columnwidth] {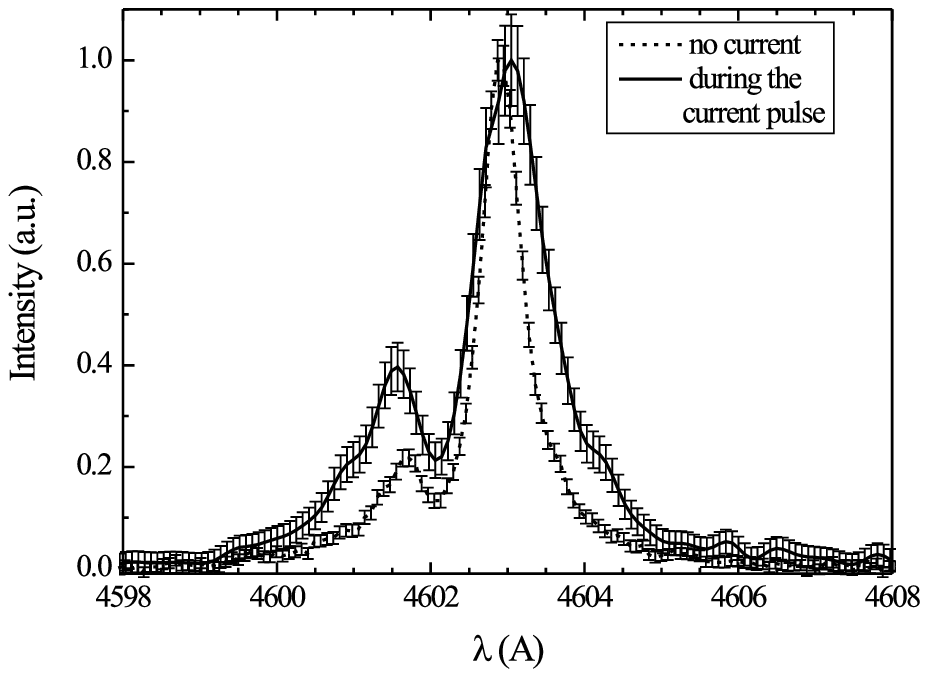}
 \caption{Comparison of the normalized spectral profiles prior and during the
          current pulse application. The measurements are performed at point
          \textbf{2}. The delay between the current start and the beginning of
          the measurement is $40~\text{ns}$, corresponding to time of maximum E-field at point \textbf{2}.}
 \label{FB_POS_profiles}
\end{figure}

Figure~\ref{FB_POS_profiles} shows a comparison between the
spectral profiles obtained at point {\bf 2} prior and after the
current-pulse application. It is clearly seen that the lines
broaden during the current conduction in the plasma, with the
forbidden component rises in intensity relative to the allowed
line and becomes shifted to shorter wavelengths. This demonstrates
the rise of the E-field in the plasma during the current
conduction. The recorded profiles are analyzed in the manner
described in Sec.~\ref{method}, taking into account the effects of
the plasma microfields, the Doppler broadening, and the
instrumental broadening. This analysis yields the electric field
amplitude as a function of time.

\begin{figure}
 \includegraphics*[width=\columnwidth] {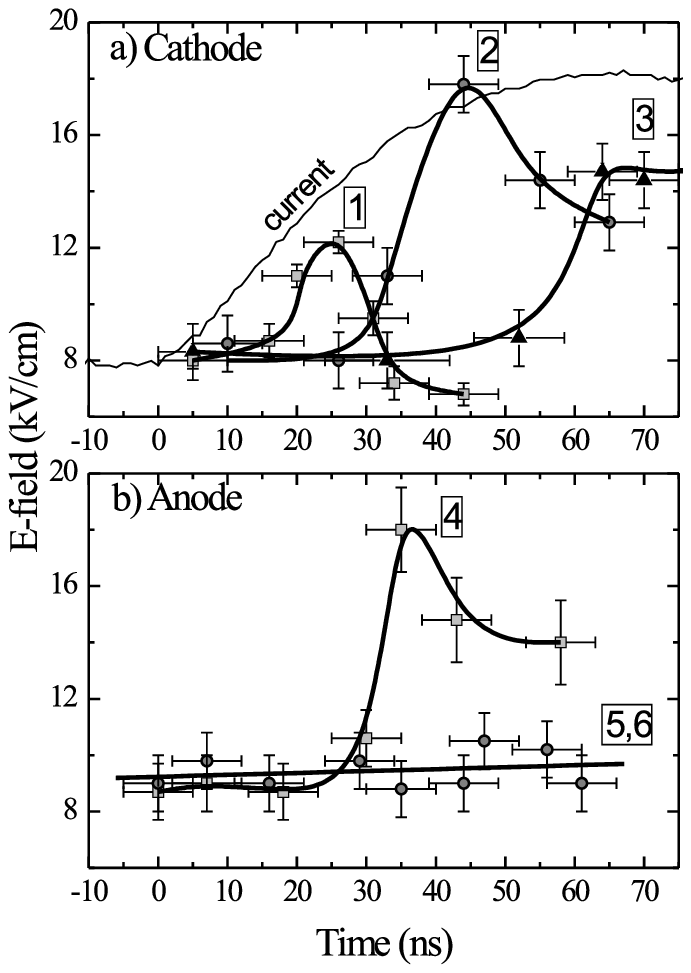}
 \caption{The evolution of the E-field obtained in the vicinity of the cathode and the anode (including the background field). The term $\textit{t}=0$ is the start of the current pulse.}
 \label{E-field_anodeVScathode}
\end{figure}

The inferred evolution of the E-fields at the 3 axial positions
near the cathode and the anode are shown in
Fig.~\ref{E-field_anodeVScathode}. The error-bars in the figure
reflect the statistical errors due to the shot-to-shot
irreproducibility. In addition, a systematic error of up to 10\%
can be expected due to the finite accuracy of the line-shape
modeling. For clarity, these have been omitted from the figure.
Thus, the absolute value of the measured E-field has an additional
error of 10\%. The rise of the E-field in the vicinity of the
cathode (Fig.~\ref{E-field_anodeVScathode}a) is observed to be
delayed between the different axial positions. This allows for
determining the axial propagation velocity of the electric field,
which is found to be $(1.5\pm 0.5)\times 10^8~\text{cm/s}$. The
velocity is found to be approximately the same between points
\textbf{1} -- \textbf{2} and \textbf{2} -- \textbf{3}, therefore,
the E-field propagation velocity is concluded to be constant over
the entire pulse. Based on the 10-ns temporal resolution, the rise
time of the E-field is inferred  to be less than $10~\text{ns}$.
The product of the field rise-time and the propagation velocity
gives an estimate of the axial scale of the changes of the
electric field. This scale of $1-2~\text{cm}$ could possibly be
related to the width of the current channel, which in previous
experiments was found to exhibit a similar scale, $\approx
1$~cm~\cite{PRL}.

In the vicinity of the anode one might expect somewhat lower E-field
values. This is plausible due to the lower B-field near the anode
and the higher fraction of heavy ions, since the latter leads to a higher
electron density following the reflection of the protons by the
propagating magnetic front. However, the E-field at point \textbf{4}
was found to have an amplitude similar to the maximum obtained near
the cathode, whereas at points \textbf{5} and \textbf{6} no rise in
the field is observed within the measurement accuracy (see
Fig.~\ref{E-field_anodeVScathode}b). The E-field in point \textbf{4}
rises at a delay compared to point \textbf{1}, demonstrating a
slower propagation of the E-field near the anode. Thus, when the
current channel near the cathode reaches the vacuum section near the
shorted end of the transmission line and as a result ceases
to flow through the plasma, the current near the anode has still not
reached point \textbf{5}, and no E-field is generated at points
\textbf{5} and \textbf{6}. An upper limit for the field propagation
velocity near the anode can thus be estimated by dividing the
distance between point \textbf{4} and point \textbf{5} by the time
interval between the field arrival at point \textbf{4} and the field
arrival to the plasma-vacuum boundary (at the shorted-end side).
This gives a  velocity of $\simeq 7\times 10^{7}~\text{cm/s}$.

We now estimate the net
rise in the E-field due to the application of the current pulse, by subtracting
from the measured total E-field ($E_{tot}$) the background field
$E_{bg}$, found in the plasma prefill prior to the current
application (see Sec.~\ref{sec_FB}). A simple subtraction of this background from the
measured E-field is not always correct. In order to demonstrate
this fact, let us assume that some external quasi-static field $E_{ex}$
at a certain direction (e.g., the Hall field) is formed in the
plasma in addition to the already present randomly directed,
fluctuating background field. On a spatial scale smaller than the
typical scale of the fluctuations, where the background field has
a distinguished orientation, the background and the external
fields should be summed up as vectors. Then, the sum of the fields
must be averaged over all possible directions of the background
field, since in the measurements the field of view of the optical
system is assumed to be larger than the fluctuation scale. As a result,
the mean value of the total measured field is given by
$\displaystyle \bar E_{tot}=\sqrt{E_{ex}^2+E_{bg}^2}$.

Table~\ref{E-res} summarizes the results of the E-field
measurements. The first column presents the position of the
measurement. The other columns give the total measured E-field and
the field obtained after the subtraction of the background field of
$8.0 \pm 1.0~\text{kV/cm}$ near the cathode and $9.0 \pm
1.0~\text{kV/cm}$ near the anode. The Table also gives the
inferred axial propagation velocity of the field ($V_{E}$).

\begin{table}[!htb]
\caption{Summary of the peak E-field parameters.}
\label{E-res}%
\begin{tabular}{|c|c|c|c|}
\hline Position &\multicolumn{2}{|c|}{Peak E-field
($\text{kV/cm}$)} & $V_{E}~(\text{cm/s})$ \\ \hline ~& Total
& Without $E_{bg}$ &~\\
\textbf{Cathode} & ~ & ~ & ~ \\
z = 1~\text{cm} (pos. 1) & $12.0\pm 0.5$ & $9.0\pm 1.1$ & ~ \\
z = 4.5~\text{cm} (pos. 2) & $17.8\pm 1.0$ & $16.0\pm 1.4$ & $1.5\pm 0.5\times 10^8$ \\
z = 8~\text{cm} (pos. 3) & $14.5\pm 1.0$ & $12.1\pm 1.4$ & ~ \\
\hline
\textbf{Anode} & ~ & ~ & ~\\
z = 1~\text{cm} (pos. 4) & $18.0\pm 1.8$ & $15.5\pm 2.0$ & $<7.0 \times 10^7$ \\
z = 4.5~\text{cm} (pos. 5) & $9.0\pm 1.0$ & 0 & ~ \\
z = 8~\text{cm} (pos. 6) & $9.0\pm 1.0$ & 0 & ~ \\
\hline
\end{tabular}%
\end{table}

\section{Discussion}
In the following, we demonstrate that the time-resolved
measurements of the electric field in several locations in the
plasma lead to two main results. The first is that
the electric fields observed are probably the Hall electric
fields resulting from the current flow in the plasma,
and the other result is that turbulent electric fields observed in
the plasma prefill prior to the current application may provide
the anomalous collisionality required for the previously observed
broadening of the current channel.

\subsection{E-field generation due to the Hall effect}
Magnetic field measurements, based on Zeeman spectroscopy,
performed in a separate study~\cite{Weingarten_Beams98}, using the
same experimental setup, show that the B-field propagates faster
near the cathode than near the anode, giving rise to a
2D-structure of the B-field propagation. We attribute this
behavior to the decrease of the magnetic field intensity with
increasing radius (the anode is the outer electrode) and to the
increasing fraction of the heavy ions (carbon ions) in the plasma
prefill closer to the anode. The parameters of the magnetic field
and plasma prefill are listed in Table \ref{param_table}. It is
expected that the rise in the E-field amplitude at each point
coincides with the arrival of the current channel, namely, the
arrival of the magnetic field front at that point. Indeed, the
measured velocities of the magnetic field penetration, $1.2\pm
0.2\times 10^8~\text{cm/s}$ near the cathode and $5 \pm 2 \times
10^7$~cm/s near the anode~\cite{Weingarten_Beams98}, are
consistent with the velocities of the E- field propagation found
here, $(1.5\pm 0.5)\times 10^{8}~\text{cm/s}$ near the cathode and
approximately a factor of two slower near the anode.
\begin{table}[!htb]
\caption{The parameters of the plasma and the magnetic field in the
present experiment. The terms $n_C$ and $n_p$ are, respectively, the
densities of the carbon and proton components of the plasma, and
$Z_{\text{eff}}^{\text{C}}$ is the effective charge of the
carbon ions.}
\label{param_table}%
\begin{tabular}{|l|c|c|}
\hline
Parameter & Cathode & Anode \\[5pt] \hline
$r$ $\text({cm})$ & 1.9 & 4.45 \\[2pt] \hline
$B_{\text{peak}}$ (T) & $1.2\pm 0.2$ & $0.5\pm 0.1$ \\[2pt] \hline
$n_{e}$ $\text({cm}^{-3})$ & $(2\pm 0.5)\times 10^{14}$ & $(2\pm
0.5)\times 10^{14}$
\\[2pt] \hline
$n_p/n_C$ & $\approx 9$ & $\approx 1$ \\
\hline
$Z_{\text{eff}}^{\text{C}}$ & $\approx 3$ & $\approx 3$ \\
\hline
\end{tabular}%
\end{table}

In a quasi-neutral plasma the electric field is given by
Ohm's law:
\begin{equation}\label{Ohms_law}
 \vec E=\frac{\vec j \times \vec B}{cn_{e}e}-\frac{1}{c}\vec
 v_{i} \times \vec B + \frac{1}{\sigma }\;\vec j,
\end{equation}
where $B$ is the magnetic field, $j$ is the current density, $v_i$
is the ion velocity, $\sigma$ is the plasma conductivity, and $c$ is
the speed of light in vacuum. Here, the electron inertia and
pressure are neglected.  We now show that the measured electric
field in the vicinity of the cathode is close to the electric field
calculated by this expression when only the first (Hall) term on the
right-hand-side is considered. The second term,
which results from ion pushing~\cite{Huba} is shown to be smaller
than the Hall term. The contribution of the third term is discussed
in Sec.~\ref{Anomalous}. For evaluating the Hall term and comparing it
to the present electric field measurements we use previous
measurements of the magnetic field and
density~\cite{PRL,Weingarten_Beams98}.

Although the B-field measurements are highly informative, they do not allow for the
reconstruction of the spatial distribution of the B-field
due to the small number of the measurement positions in the A-K gap.
Nevertheless, these measurements demonstrate a strong
correlation between the rise of the magnetic field at each measurement position
and a drop of the electron density (see, e.g., \cite{PRL}).
We note that the prominent density drop is closely related to the
phenomenon of ion-species separation~\cite{PRL}, mentioned in
Sec.~\ref{introduction}, in  which the magnetic field only
penetrates into the heavier component of the plasma (carbon ions)
while the light ions (protons) are reflected ahead by the magnetic
field front. In Ref.~\cite{Weingarten_Beams98} a map of the electron
density distribution was measured at different instants of the
current conduction in an experimental
setup similar to the present one.
These data show that the density drop propagates
axially, exhibiting different propagation velocities at different
radial positions. Thus, assuming the density drop in the A-K gap is
correlated with the arrival of the magnetic field front everywhere,
the measured evolution of the density map can be used to determine
the distribution of the B-field front.

\begin{figure}
 \includegraphics*[width=\columnwidth] {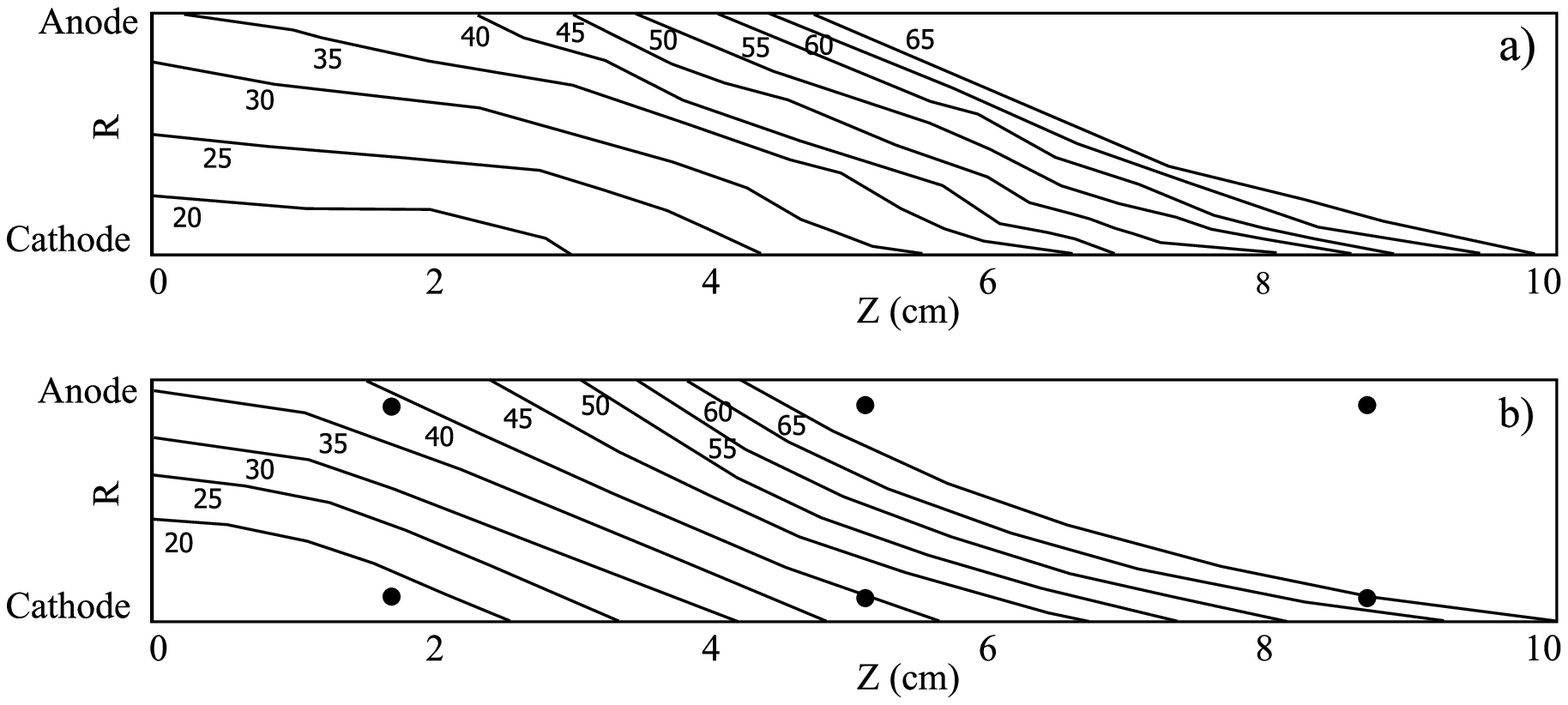}
 \caption{The map of the magnetic field propagation a) obtained from the
          electron density evolution (contours denote the location at which the
          density starts to drop), b) generated using Eq.
          (\protect\ref{bmap_fun}) (here the contours are the locations where the
          magnetic field is 0.2~T). The numbers on the plot represent the time
          in nanoseconds relative to the beginning of the current pulse. The solid circles in (b) represent the positions of the present E-field measurements.}
 \label{B-front}
\end{figure}

We now fit an analytical expression for the magnetic field temporal
distribution. Based on the previous
measurements~\cite{PRL,B-fieldRon} we approximate the B-field axial
distribution at each radial position by a step-like function of a
front width $\delta$, i.e., the B-field rises within a distance
$\delta$ that  corresponds to the current-channel width. This step
front propagates axially with a constant velocity. The propagation
velocity is lower at larger radii, similarly to the
density drop~\cite{Weingarten_Beams98}. At the plasma boundary,
$z=0$, the magnetic-field temporal behavior was found to be similar
to that of the generator current, approximated by
$I_{up}=I_{max}\sin \left( \frac{ \pi t}{2\tau }\right)$, where
$\tau$ is the time corresponding to the peak-current. Therefore, we
assume that the amplitude of the field in the back of the step-front
(which is the height of the step) also rises in time as
$sin\left(\frac{ \pi t}{2\tau }\right)$. For the radial distribution
of the magnetic field we assume that the field at the rear of the field-front
decreases towards the anode as $1/r$, due to the
cylindrical geometry of the experiment. Under these assumptions we
find that the magnetic field can be approximated by:
\begin{eqnarray}
\label{bmap_fun}
&\displaystyle B(r,z,t)=B_{0}\frac{r_{c}}{r}\sin \left( \frac{\pi t}{2\tau }%
\right)\times &  \nonumber \\
&\displaystyle \times \frac{1 }{2}\left[ 1-\tanh \left( 2\,\frac{%
\displaystyle z-v_{c}\frac{r_{c}}{r}t}{\delta }\right) \right],
\end{eqnarray}
where $B_{0}$ is the peak magnetic field, $r_{c}$ and $r_{a}$ are
the cathode and anode radii, $0<t<\theta$ (here we focus on the
period of the current flow through the plasma lasting until
$\theta=70~\text{ns}$ that denotes the arrival of the current
channel at the plasma transmission-line boundary), $v_{c}$ is the
B-field propagation velocity near the cathode, and $\delta $ is the
current-channel width. In our experiment $\tau\approx\theta$. The
variables $r$ and $z$ are limited by $r_{c}<r<r_{a}$ and
$0<z<z_{end}$, where $z_{end}$ is the position of the plasma
transmission-line boundary. In the calculations, the current channel
width $\delta$, which is assumed to be broadened by diffusion,
varies in time as $\delta=\lambda_{e}+\delta_{\theta}t/\theta$,
where $\lambda_{e}$ is the electron skin depth. The term
$\lambda_{e}+\delta_{\theta}$ is the front-width at the instant
$t=\theta$. Since at time $t=\theta$ $\delta$ is measured to be
about $1~\text{cm}$, which is much larger than $\lambda_{e}$, we
obtain $\delta_{\theta}\approx\delta(t=\theta)\approx1~\text{cm}$.
The comparison of the map of the B-front propagation determined from
the previous measurements of the electron density drop to the map
generated using the semi-analytical formula (2) is shown in
Fig.~\ref{B-front}, demonstrating a good agreement between the two
approaches.

Using the analytic form of the magnetic field and the fact that $\vec B =
B \vec e_{\varphi}$, the Hall electric field is expressed by:
\begin{eqnarray}
 \label{Eh}
&\displaystyle \vec E_{Hall} = \frac{[rot \vec B \times \vec B]}{%
4 \pi\: n_{e}\: e}=&  \nonumber \\
&~&  \nonumber \\
&\displaystyle =\frac{1}{4\pi\: n_{e}\: e}\left[\vec e_r\left(B\:\frac{1%
}{r}\frac{\partial (rB)}{\partial r}\right)+\vec e_z \left(B\frac{%
\partial B}{\partial z}\right)\right]&.
\end{eqnarray}
Finally, by inserting Eq.~(\ref{bmap_fun}) into Eq.~(\ref{Eh}),
the axial ($E_{Hall,z}$) and radial ($E_{Hall,r}$)
components of the electric field are found to be:
\begin{eqnarray}
&\displaystyle E_{Hall,z}=\frac{B_{0}^{2}\:r_{c}^{2}\:\sin ^{2}\left( \frac{%
\displaystyle \pi t}{2\tau }\right) }{8\pi \:n_{e}\:e\:r^{2}\:\delta }\:%
\left[ 1-\tanh (\zeta) \right] \times  \nonumber  \label{Er} \\
{~} &\times \left[ 1-\tanh ^{2}(\zeta) \right],& \\
&~&  \nonumber \\
&\displaystyle E_{Hall,r}=\frac{ B_{0}^{2}\:r_{c}^{3}\:v_{c}\:t\:\sin
^{2}\left( \frac{\displaystyle\pi t}{2\tau }\right) }{8\pi
\:n_{e}\:e\:r^{4}\:\delta }\:\left[ 1-\tanh (\zeta) \right] \times  \nonumber
\\
&\times \left[ 1-\tanh ^{2}(\zeta) \right] ,&  \label{Ez}
\end{eqnarray}
where $\displaystyle\zeta=2\,\frac{\displaystyle
z-v_{c}\frac{r_{c}}{r}t}{\delta}$.

We now compare the magnitude of the measured E-field with the
calculated $E_{Hall}$ using the expressions (\ref{Er}) and
(\ref{Ez}). We note that for this comparison, knowledge of the
electron density is essential. Here, again, we use the previous
measurements by Weingarten {\it et al.}~\cite{Weingarten_Beams98}. We
assume that the magnetic field reflecting the protonic component of
the plasma and thereby causing the density to drop, penetrates only
into the carbon plasma~\cite{PRL}. The electron density of the
carbon plasma (penetrated by the magnetic field) is determined here
for each position at the time of the arrival of the magnetic front
at the transmission-line boundary of the plasma. At this time, at
the generator side of the plasma, namely at point \textbf{1}, the
electron density drops from $2\times 10^{14}~\text{cm}^{-3}$ to
$5\times 10^{13}~\text{cm}^{-3}$. In the middle of the plasma (at
point \textbf{2}) the density only drops to
$10^{14}~\text{cm}^{-3}$, and at the transmission-line boundary of
the plasma (at point \textbf{3}) no density drop is found, rather a
slight rise to $\approx 2.5\times 10^{14}~\text{cm}^{-3}$ is
observed. Using these data, the evolution of $n_{e}$ is incorporated
into the calculations of $E_{Hall}$ by synchronizing the change of
the local value of the electron density with the rise of the local
magnetic field, namely $\displaystyle n_{e}\left( z,r,t\right)
=n_{0}-\Delta n\left( z,r\right) \frac{B\left( z,r,t\right)
}{B_{max}\left( r,t\right) }$. Here, $\Delta n\left( z,r\right)$
denotes the density drop at the particular position for which the
electric field evolution is calculated, and $\displaystyle
B_{max}=B_{0}\frac{r_{c}}{r}\sin \left( \frac{\pi t}{2\tau}\right)$.

\begin{figure}
 \includegraphics*[width=\columnwidth] {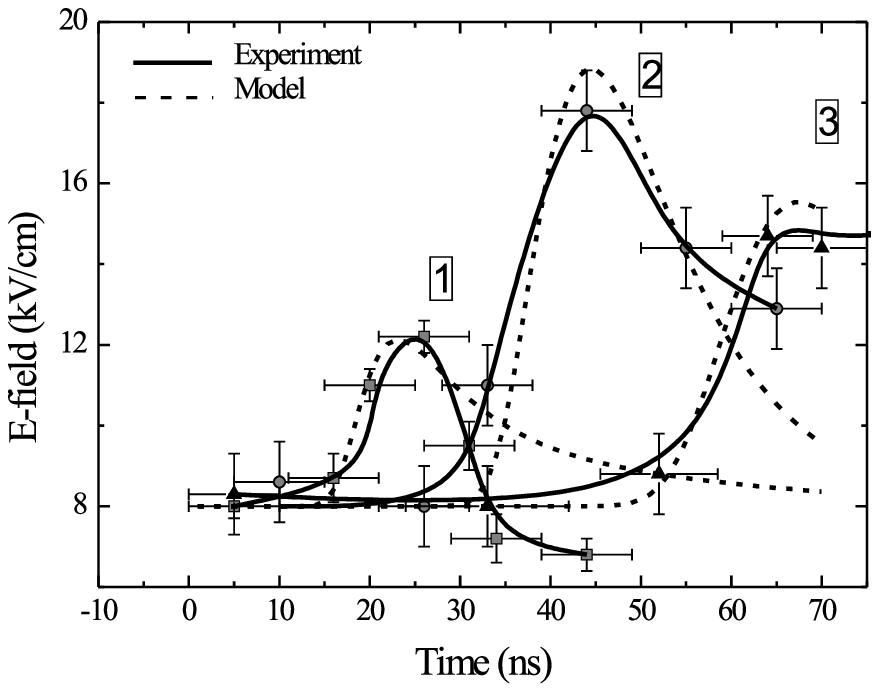}
 \caption{Comparison of the electric field evolution observed in the
          experiments with the Hall-electric field calculated using the
          semi-analytical approximation. The calculated total Hall-field is a vector sum
          of $E_{Hall,z}$ and $E_{Hall,r}$. The background E-field of 8~kV/cm,
          found in the flash-board plasma,
          is added to the calculated values in the manner described in Sec.~\ref{sec_FB}.}
 \label{Evsmodel}
\end{figure}
Thus, the evolution of the Hall electric field is calculated for the
three different axial positions at which the electric field
measurements were performed. Figure~\ref{Evsmodel} gives a comparison
between the calculation results and the experimental data near the
cathode. The Hall field presented is the magnitude of the electric
field including its axial and radial components, to which the
$8~\text{kV/cm}$ turbulent background has been added according to
$E=\sqrt{E^2_r+E^2_z+E^2_{bg}}$ (see Sec.~\ref{sec_FB}). It is seen
that the calculated Hall-field fits well the measured values, except
for point \textbf{1} at relatively late times. The relatively large
deviation seen for point \textbf{1} can possibly be explained by the
more pronounced $n_{e}$ drop observed at this point, which may
reduce the turbulent background E-field, contrary to our assumption
in the model that the background field is constant.

Such a good agreement between the experimentally measured electric
field and the calculated Hall electric field, based on the B-field
and density measurements, as shown in Fig.~\ref{Evsmodel}, suggests
that the observed E-field is most likely the Hall field, generated
by the current. Although a thorough understanding of the field
evolution requires a self-consistent modeling that, for example,
would also consider the ion reflection and species separation
observed in the experiment, this agreement indicates that the
Hall field could play a crucial role in the magnetic field
penetration.

The model discussed here predicts the E-field near the anode
(point~\textbf{4}) to be a factor of 4 lower than that near the
cathode at point~\textbf{1} (B-field drops by a factor of 2 near the
anode, see Table \ref{param_table}), in disagreement with the
measurements (see Fig. \ref{E-field_anodeVScathode}). A possible
explanation is a rise of the turbulence near the anode during the
current flow. In order to test this explanation, one could measure
the axial velocities of the accelerated ions both near the anode and
near the cathode. Since the ions accelerate axially mostly due to
the Hall-field, if indeed the electric field is mostly turbulent
near the anode then the axial ion velocities near the anode will be
substantially lower than they are near the cathode, but the velocity
distribution will be much broader.

The contribution of the ion motion to the electric field in Ohm's
law (the second term of the RHS of Eq. (\ref{Ohms_law})) is small.
This contribution can be estimated by comparing $\left\vert
\vec {j}/en\right\vert $ to $\left\vert
\vec {v_{i}}\right\vert$. Approximating the magnitude of the
current density by $\left\vert \vec {j}\right\vert \simeq
cB/4\pi \delta $, using the measured ion velocities from
Ref.~\cite{PRL}, and assuming the ions are accelerated in the
direction perpendicular to the B-field front, we find the
contribution of the convective term in Eq.~(\ref{Ohms_law}) to be
much smaller than the contribution of the Hall term. If instead of
using the measured ion velocities we use the MHD velocity
$\left\vert \vec {v_{i}}\right\vert \simeq B/ \sqrt{4\pi \rho }$
(where $\rho$ is the mass density ahead of the current front) as the
maximal ion velocity, we find that the contribution of the Hall term
is larger than that of the ion motion as long as $\delta <\left(
c/en\right) \sqrt{\rho /4\pi}$ (the effective ion skin depth). Here,
$\rho \cong 6.7\times 10^{-10}~\text{g/cm}^3$ yields an effective
ion skin depth of $\cong 2.3~\text{cm}$ that is much larger than
$\delta \cong 0.5~\text{cm}$ (the current-channel width observed in
the experiments), thus the ion term is expected to be smaller.

\subsection{Anomalous collisionality}
\label{Anomalous}
As suggested in Sec.~\ref{sec_FB}, the background electric field may be
attributed to an ion-acoustic instability in the plasma. In this
case, this electric field should cause a relatively strong electron
collisionality in the plasma prior to the application of the current. Here
we describe the effect of such a strong collisionality on the magnetic-field evolution and
the current distribution.

Following Ref.~\cite{Weingarten_turbulence}, the effective collision
frequency $\nu^{tur} _{ei}$ associated with the ion-acoustic turbulent electric fields in the
plasma is:
\[
\nu^{tur} _{ei}\simeq \omega _{pe}\frac{\langle E^{2}\rangle }{8\pi n_{e}kT_{e}}.
\]
Assuming that the turbulent E-field is of the order of the
background E-field of $8~\text{kV/cm}$ measured prior to the current
application, and using $n_{e}=5\times 10^{13}~\text{cm}^{-3}$ and
$T_{e}=10~\text{eV}$ ($10~\text{eV}$ being the upper limit of the electron temperature in the plasma prefill),  the expected collision frequency is $\nu^{tur}
_{ei}\simeq 10^{10}~\text{Hz}$. This frequency is much higher than
Spitzer's collision frequency, which for the above parameters is
$2\times 10^{8}~\text{Hz}$. Such a high collision frequency is
expected to increase the rate of magnetic field penetration into the
plasma and also to broaden the current channel.

In spite of the relatively high anomalous collision frequency
suggested here, it is still insufficient to explain the fast B-field
penetration by diffusion that requires collision frequency of about
$7\times 10^{11}~\text{Hz}$ ~\cite{B-fieldRon}. In addition,
step-like spatial profile of the magnetic field is inconsistent with
the  exponential decay predicted by diffusion. A different
mechanism, such as the Hall-field induced mechanism
\cite{Gordeev,Fruchtman}, is thus expected to cause the fast
magnetic field penetration. Still, the high anomalous collision
frequency may help to explain the observed broad current-channel,
not understood thus far. According to
Ref.~\cite{Weingarten_turbulence}, in a steady Hall-induced
penetration the current-channel width is estimated to be:
\[
\delta\simeq L~\frac{\nu _{ei}}{\omega _{pe}}\frac{c}{V_{Ae}},
\]%
where $V_{Ae}$ is the Alfv\'{e}n velocity of electrons,
$\displaystyle V_{Ae}\equiv B/(4\pi n_{e}m_{e})^{1/2}$ ($m_{e}$ is
the electron mass), and $L$ is the inter-electrode spacing. Although
in the present experiment the plasma and field dynamics do not reach
a steady state, we use this relation to obtain an estimate for the
current-channel width, assuming Hall-induced penetration. For
$L=2.5~\text{cm}$ and the anomalous collision frequency estimated
above, we find that $\delta\simeq 5~\text{mm}$. This value is an
order of magnitude larger than the electron skin-depth $c/\omega
_{pe} \sim 0.5~\text{mm}$ and consistent with the measurements
~\cite{Weingarten_Beams98} used for our model
($\delta(t=\tau)=1~\text{cm}$). Thus, while the anomalous collision
frequency estimated from the turbulent electric fields is
insufficient to explain the rate of magnetic field penetration, it
does explain reasonably well the current channel width.
We note that in a previous investigation~\cite{Weingarten_turbulence}
it was concluded otherwise, based on an estimated upper level of a
$5~\text{kV/cm}$ electric field.
However, in that experiment the plasma source was a plasma gun rather than a surface flashover as
in the present work. Also, the previous E-field measurements might have been related to
a secondary plasma, since they were based on hydrogen line-emission
that also originated at electrode sputtering~\cite{Arad_flashboard}.

The anomalous collision frequency predicted above is about three times higher
than the minimal collision frequency required for a collisional current channel dominated
by the Hall-field penetration~\cite{FruchtmanPRA}, $\nu _{min}=B/\left[ L(8\pi
m_{e}n_{e})^{1/2}\right]$, which is estimated to be $3\times 10^{9}~\text{Hz}.$
This further supports the possibility of a Hall-field induced magnetic-field
penetration.

We note that for $B<0.3~\text{T}$ the value of $\nu^{tur} _{ei}$ is
larger than the electron gyrofrequency,  meaning there is no
electron magnetization and the resistive term (the third term on the
RHS of Eq.~(\ref{Ohms_law})) is thus dominant in determining the
electric field at the beginning of the current conduction.
Therefore, the Hall mechanism is not applicable in these conditions.
The collisionality, though rather high, also cannot explain the
velocity of the magnetic field penetration at this early stage of
the pulse and the question regarding the mechanism of the initial
magnetic field penetration remains open.

\section{Summary}
For the first time the electric fields in current-carrying plasmas
are measured with high spatial and temporal resolutions. This is
achieved by a diagnostic method that is based on line-shape
analysis of dipole-forbidden transitions combined with laser
spectroscopy. The measured electric fields exhibit a good
agreement with the Hall field calculated using measurements of the
time-dependent magnetic-field and density distributions in the
current-carrying plasma. These results indicate that in the
present experiment the Hall mechanism can be dominant in the
magnetic field penetration into the plasma. Moreover, it is
noteworthy that the modeled Hall field, which depends linearly on
$1/n_{e}$, reproduces well the present E-field measurements,
particularly in light of the sharp density drop observed due to
the ion-species separation phenomenon. It is important to note,
however, that in order to further conclude that the Hall electric
field induces the magnetic field penetration, as suggested in
Refs.~\cite{Gordeev,Fruchtman,FruchtmanPRA}, it is yet to be shown
that the evolution of the electric fields must be consistent with
the magnetic field propagation in accordance with Faraday's law:
\begin{equation}
\frac{\partial \mathbf{B}}{\partial t}=-\frac{1}{c}\mathbf{\nabla
\times E} \;  \label{Bevolution}
\end{equation}

A rather high level of the electric field ($\approx 8$~kV/cm) is
observed in the plasma prefill prior to the pulsed current
application. Such electric fields, if present in the entire prefill plasma,
may give rise to an anomalous
collisionality that is consistent with the Hall mechanism and may
help explaining the observed current channel width. A
clarification of this effect and the nature of this possible
turbulence can be a subject of future investigations.

\begin{acknowledgments}

The authors thank R. Arad for useful comments and are grateful to Yu. V.
Ralchenko for his help in the collisional-radiative calculations. Highly
valuable discussions with H.-J. Kunze and H. R. Griem are appreciated. We thank
P. Meiri for his skilled technical assistance.

This work was supported in part by the German-Israeli Project Cooperation
foundation (DIP), the U.S.-Israel Binational Science Foundation (BSF), the Israel Science Foundation (ISF),
and by Sandia National Laboratory (USA).
\end{acknowledgments}

\end{document}